\documentclass[aps,pre,twocolumn]{revtex4}
\usepackage{ae}
\begin{document}  
%\maketitle
\narrowtext

\noindent
{\bf Comment on: Competing Interactions, the Renormalization Group, 
and the Isotropic-Nematic Phase Transition}

In a recent letter Barci and Stariolo (BS)\cite{BaSt07} generalized the well known
Brazovskii model to include an additional rotationally invariant 
quartic interaction and study this model in two dimensions (d=2). 
Authors do not give any specific example
of a  microscopic system to which this generalized 
Landau-Ginzburg-Wilson (LGW) action corresponds.
After a brief discussion of a possible renormalization
group (RG) treatment, they proceed to study their model using a mean-field
Hartree approximation. They then find that
when $u_2>0$ the model exhibits striped (lamellar) phase, but when
$u_2<0$ it shows a nematic order.

Brazovskii model is notoriously difficult to renormalize. 
There exist, however, 
well known lattice models which exhibit exactly the same
phenomenology and are much more susceptible to the 
RG treatment.  
One example is Widom's isotropically spatially 
frustrated lattice model of microemulsion~\cite{Wi86}.  The model exhibits
isotropic-to-lamellar phase transitions, which can be
studied by mapping it directly onto anisotropic  $O(6)$ field 
theory~\cite{LeDa90}.  
This field theory has the lower critical dimension $d_l=2$ and
the upper critical dimension $d_u=4$ (away from the isotropic Lifshitz point
for which $d_l=4$ and $d_u=8$).  
%The RG treatment shows that the
%initial conditions --- determined by the microscopic Hamiltonian --- 
%place the RG flow in the region of the parameter space where no
%stable fixed point is available so that the phase transition becomes
%fluctuation induced first order, in agreement with the Brazovskii  analysis.
%We stress that the lower critical dimension for the effective field theory 
%in this case (away from
%the Lifshitz point) is $d_l=2$.  
Since existence of a lattice 
diminishes the role of fluctuations, we  expect that the lower critical
dimension for the field theory considered in Ref. \cite{BaSt07} should be
$d_l > 2$.  Below the lower critical dimension no mean-field
theory can be trusted {\it even qualitatively}.  Since the 
Hartree approximation
used in Ref. \cite{BaSt07} is nothing more than a mean-field theory it
is bound to fail in 2d. 
%In fact, one can easily derive the Hartree approximation by applying 
%the usual Gibbs-Bogoliubov 
%inequality --- with a Gaussian variational action --- to the original
%LGW action.  In this respect it is difficult to judge {\it a priory}
%any advantage of the Hartree approximation as compared to, say, 
%applying the same 
%Gibbs-Bogoliubov inequality directly to the microscopic model without first
%going through a Hubbard-Stratanovich transformation to map it onto 
%an effective field
%theory which is then expanded to only two leading orders.  
In this Comment I will argue that the situation of the 
theory of Ref. \cite{BaSt07} is even more difficult, since the 
lower critical dimension for this model is actually $d_l=3$.

Let us first consider the lamellar phase found when $u_2>0$. Suppose
that the symmetry is broken in such a way that the lamellae are parallel
to the x-axis  (parallel to the x-y plane in 3d). 
We want to study the fluctuations of the interfaces separating
the high and the low order parameter states.  
Suppose that one of the interfaces  
lies along the x-axis (is in the x-y plane in 3d).  
At finite temperature this interface will
fluctuate.  We want to find an effective Hamiltonian which controls
these fluctuations.  Clearly, this Hamiltonian must be invariant under the
transformation $h \rightarrow -h$, where $h$ is the height of the 
interface over the projection plane.    
Furthermore,  since the original 
LGW action of  Ref. \cite{BaSt07} is invariant under 
arbitrary translations and rotations, the effective
interfacial Hamiltonian must be invariant under 
$h \rightarrow h+a+{\bf b}\cdot {\bf x}$, where ${\bf x}$ is an arbitrary
vector in the projection plane of the lamella, 
and $a$ and ${\bf b}$ are arbitrary
constants. To leading order in $h$, 
the interfacial Hamiltonian must, therefore, be 
${\cal H}=\frac{\kappa}{2}\int d^{d-1}{\bf x} (\nabla^2 h)^2 $, where 
$\kappa$ is an effective bending modulus. 
%Due to the underlying rotational
%invariance this Hamiltonian is much softer than the one for, say, liquid
%vapor interface, dominated by the surface tension.  
We now study
the fluctuations of these interfaces.  Define a local
width of an interface $w$ 
as $w^2=\langle [h({\lambda_0/2)})-h(0)]^2 \rangle$, where $\lambda_0$
is the wavelength, $\lambda_0=2 \pi/k_0$, of the order 
parameter in the symmetry
broken phase. It is then easily found that 
for a two dimensional system (1d
interfaces) the interfacial width $w$ diverges as 
$w^2 \sim T \lambda_0^2 L/\kappa$, where $L$ is the system size.  
Thus, lamellar order is impossible
in 2d.  In 3d (2d interfaces), $w$ diverges as 
$w^2\sim (T \lambda_0^2/\kappa)\ln(L/\lambda_0)$, allowing for
a pseudo-long-range lamellar order.  This fact was already known
to Landau in the $1940$'s 
and is commented in the original Brazovskii paper,
who also finds logarithmic divergences  beyond the
Hartree approximation. Note that existence of 
a crystalline substrate can stabilize a lamellar structure in 2d by 
breaking the rotational symmetry. No such terms, however, are considered 
in the LGW action analyzed in Ref.\cite{BaSt07}. 
It is, then, clear that no mean-field theory (Hartree included) 
can be used to study the Brazovskii type models in 2d.  
%The fluctuations will destroy completely the lamellar (stripe) order found in 
%Ref.\cite{BaSt07}. 

For $u_2<0$, BS find an isotropic-nematic transition 
transition with {\it mean-field exponents}.  They then speculate that fluctuations
will turn this transition into the Kosterlitz-Thouless (KT) one.  
Nowhere is this statement proven explicitly and, indeed, no such proof is possible.
In 2d the fluctuations can completely destroy
a mean-field phase transition, as happened for the
lamellar phase discussed above.
The fluctuations can also change the
nature of the phase transition to something very different and non-universal.   
Simply because the
coarse-grained LGW action has a nematic symmetry, does not imply that a microscopic 
(fine-grained) model will have a KT transition.  There is no such strong universality
in 2d!  For example, there is a class of generalized XY models all with the same
LGW action, but whose critical behavior depends on the precise form of the 
microscopic interaction potential~\cite{Vi07}. 
To conclude, the isotropic-to-lamellar phase transition found by
BS can not exist in 2d.  As far as the isotropic-nematic transition, 
nothing about its order or its universality class can be said based on 
the coarse-grained LGW action presented in Ref.\cite{BaSt07}.
%To conclude,
%in 2d, mean-field analysis is not sufficient to predict existence of
%a phase transition, and symmetry arguments alone are not
%sufficient to account for its universality class or even its order.

This work was supported in part by the CNPq.\\
\noindent
Yan Levin \\
Instituto de Física, UFRGS\\ CP 15051, 91501-970, Porto Alegre, RS,
Brazil

\end{document}